\documentclass{iaus}
\usepackage{graphicx}

\title[Abundances in the Bulge and Galactic Center] 
{Abundance Patterns in Stars in the Bulge and Galactic Center}

\author[Cunha et al.]   
{Katia Cunha$^1$%
  \thanks{On leave from Observat\'orio Nacional; Rio de Janeiro - Brazil},
 V. V. Smith$^1$; K. Sellgren$^2$; R. D. Blum$^3$; \break S. V. Ram\'{\i}rez$^4$ \and D. M. Terndrup$^2$
 }

\affiliation{$^1$NOAO, Casilla 603, La Serena, Chile 
\break email: kcunha@noao.edu; vsmith@noao.edu\\[\affilskip]
$^2$The Ohio State University, 140 West 18th Ave., Columbus, OH 43210,
USA; \break email: sellgren@astronomy.ohio-state.edu; terndrup@astronomy.ohio-state.edu\\[\affilskip]
$^3$NOAO, 950 N. Cherry Ave., Tucson, AZ 85219, USA; email: rblum@noao.edu\\[\affilskip]
$^4$Infrared Processing and Analysis Center, Caltech, Mail Code 100-22, Pasadena, CA 91125, USA;
email: solange@ipac.caltech.edu\\[\affilskip]}
\pubyear{2007}
\volume{245}  
\pagerange{1--4}
\date{?? and in revised form ??}
\jname{Formation and Evolution of Galaxy Bulges}
\editors{M. Bureau, et al. eds.}
\begin{document}

\maketitle

\begin{abstract}
We discuss oxygen and iron abundance patterns in K and M red-giant members of the Galactic
bulge and in the young and massive M-type stars inhabiting the very center of the Milky Way.
The abundance results from the different bulge studies in the literature, both in the
optical and the infrared, indicate that the [O/Fe]-[Fe/H] relation in the bulge does not follow
the disk relation, with [O/Fe] values falling above those of the disk.
Based on these elevated values of [O/Fe] extending to large Fe abundances, it is suggested 
that the bulge underwent a rapid chemical enrichment with perhaps a top-heavy initial mass function.
The Galactic Center stars reveal a nearly uniform and slightly elevated (relative to solar)
iron abundance for a studied sample which is composed of 10 red giants and supergiants.  Perhaps of more significance
is the fact that the young Galactic Center M-type stars show abundance patterns that are reminiscent of those
observed for the bulge population and contain enhanced abundance ratios of $\alpha$-elements 
relative to either the Sun or Milky Way disk at near-solar metallicities.

\end{abstract}

\firstsection 
\section{Introduction}

Abundance patterns in different populations in the Milky Way can shed light on Galaxy formation and 
its chemical evolution. Studies carried out over the last decade have provided accurate
abundance patterns for stars in the Milky Way disk and halo so that these populations are now fairly-well mapped.
There is significantly less information, however, on the Galactic bulge population and, in particular
the Galactic Center, due to difficulties associated with heavy extinction.

The stellar content belonging to the bulge population is old, with an age of 10--12 Gyr 
(e.g. Zoccali et al. 2003) and chemical abundance studies in both low and high-resolution 
find that the bulge is overall metal-rich with a large abundance spread that spans $\sim$1.5 dex (e.g. 
Fulbright et al. 2006). 
The Galactic Center, on the other hand, contains a significant young population, including many luminous
and massive stars. Those known to date are concentrated in three 
clusters within 60 pc of the Galactic Center: the Central Cluster, the Arches Cluster and the Quintuplet Cluster. 
The youngest stars in these clusters were formed recently with
ages ranging roughly between 1 -- 9 Myr. In this contribution we summarize the
abundance results obtained recently for the bulge and Galactic center
and briefly discuss their implications in light of Galactic chemical evolution. 

\section{Oxygen and Iron Abundances in the Bulge and Galactic Center}

The evolution of [O/Fe] as a function of metallicity is an important constraint
for models of chemical evolution as this ratio ultimately probes the relative importance
of SN II (that produce oxygen) relative to SN Ia (that are the main producers of iron). 
In the Milky Way, O/Fe versus Fe/H can be compared for various stellar samples
belonging to the different galactic populations. The overall behavior 
of [O/Fe] versus [Fe/H] in the disk and halo, in particular, is well defined and this 
behavior is illustrated in the two panels of Figure 1 (as blue open symbols). 

\subsection{The Bulge}

In the top panel of Figure 1, we also show the bulge abundance results from recent studies in the literature: 
Rich \& Origlia (2005); Cunha \& Smith (2006); Lecureur et al. (2007) and Fulbright et al. (2007). 
It is apparent from the figure that the abundances from the different bulge studies 
(represented by red symbols) generally overlap.
Such an agreement between the results in the literature is pleasing given that these studies
use different scales for the adopted stellar parameters (effective temperatures, surface gravities
and microturbulent velocities) and abundance indicators. In particular, Fulbright et al. (2007), Zocalli et al. (2006) and
Lecureur et al. (2007) analyze K giants from optical spectra, while the Rich \& Origlia (2005) and Cunha \& Smith (2006) 
analyses are in the infrared. The oxygen abundance indicators in the infrared are the OH vibrational-rotational molecular transitions 
(the molecular equilibrium is set also by the CN and CO molecules), while in the optical, the forbidden oxygen line 
at $\lambda$ 6300\AA\ is the sole abundance indicator and in metal rich stars there is the issue of contamination 
by CN molecular lines that must be accounted for. 

\begin{figure}
\centerline{
\includegraphics[height=5.in,width=3.5in,angle=0]{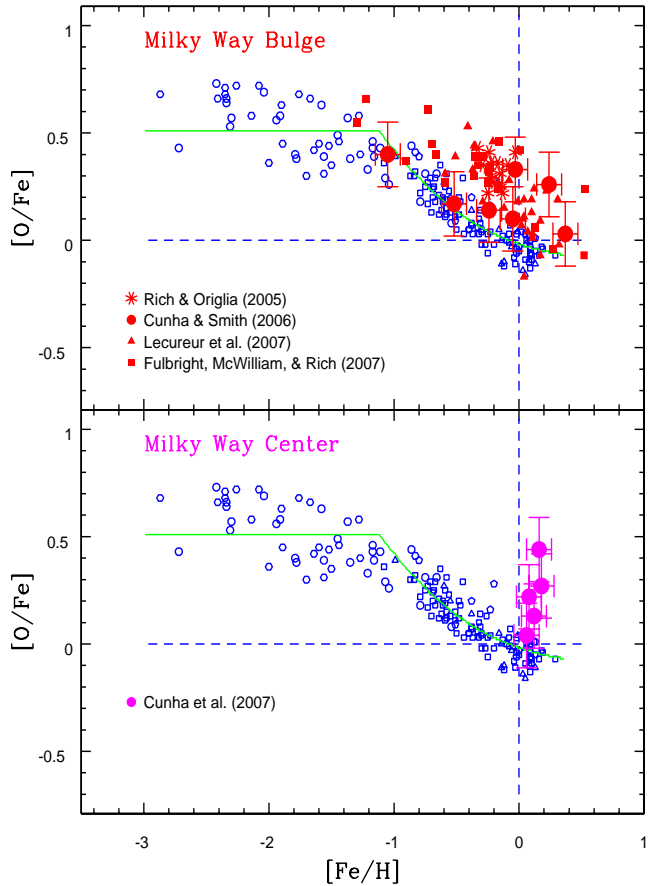}
}
  \caption{Top and bottom panels show the behavior of [O/Fe] versus [Fe/H] for the halo and the disk populations
(represented by the blue open symbols). The green solid curve represents a simple chemical evolution model for 
SN II and SN Ia yields.
The top panel also shows the abundance results for the old K and M bulge red-giants from different studies 
in the literature.
The bottom panel shows the abundance results for the young M-type giants and supergiants members of
the Central and Quintuplet clusters in the Galactic Center.
}\label{fig:oversusfe}
\end{figure}

It is of significance, therefore, that the bulge stars in all studies generally fall above the trend delineated
by the stars in the Galactic disk, indicating firmly that the bulge contains enhanced abundance ratios of
[O/Fe] relative to typical disk values.
This displacement to enhanced [O/Fe] values can be explained in terms of a rapid chemical 
enrichment in the bulge in times before SN Ia's could add their Fe content to the gas that formed the stars. 
By the time the SN Ia’s turn on the gas is already quite metal rich and these supernovae can then add their iron and bring 
the [O/Fe] values down. A top-heavy IMF may also be needed to explain the enhanced $\alpha$-element abundances
observed in the bulge. (See Ballero et al. 2007 for a more complete picture in light of a chemical evolution model.) 

\subsection{The Galactic Center}

Abundances in stars in the Galactic Center have just recently started to be determined. 
The oxygen and iron abundance patterns observed for the only Galactic Center sample studied so far are shown in the 
bottom panel of Figure 1. The abundance results are taken from Cunha et al. (2007).  
Concerning the behavior of [O/Fe] it is clear that the Galactic Center also shows oxygen enhancements
when compared to the trend observed for the disk at roughly solar metallicities: the average value of [O/Fe] for the studied
targets is elevated on average by 0.3 dex relative to the disk or the solar value. It is apparent from the figure that
the iron abundances obtained are slightly above solar with a relatively small abundance scatter.
Iron abundances in a sample of 10 Galactic Center stars were previsouly derived by Ram\'{\i}rez et al (2000).
The spread in the Fe abundances in both Ram\'{\i}rez et al. (2000) and Cunha et al. (2007) studies are similar and could be explained in terms
of the uncertainties in their respective abundance analyses. Although there is a limited number of stars studied in the Galactic
Center so far, there is a hint that the Fe abundance spread is much smaller when compared to the large metallicity spread
found for the older bulge population.

Chemical evolution models for the distinct environment in the Galactic Center, including an extended star formation 
history as well as several sources of gas outflow and inflow, are not available in the literature.
In general terms, however, the observed signature of enhanced oxygen (as a representative of $\alpha$-elements) abundances
points to the dominant role played by massive stars (that explode as SN II).
As discussed previously the Galactic Center targets analyzed are young and massive (were born ‘recently’)
and the established trend in systems with extended star formation history and ongoing star formation is that 
the [$\alpha$/Fe] should follow the trend towards scaled solar composition. This is not what is observed for the Galactic Center.

Two simple scenarios can be suggested based on the abundances observed for the Galactic Center:

1) The Galactic Center may have been dominated by SN II relative to SN Ia's
over its entire history; perhaps the progenitor binary systems that give rise to SN Ia's
do not survive in the active Galactic Center environment. 

2) The initial mass function (IMF) in the Galactic Center is more weighted towards
massive stars with low-mass star formation suppressed (Morris \& Serabyn 1996). 

It is possible that one could have the gas come from an outside reservoir, such as $\alpha$-enhanced gas ejected via mass loss from the 
population of bulge red-giants, which is happening continuously over the red-giant lifetime.
In fact, the abundance pattern observed in the Galactic Center young stars closely resembles what would be expected 
from a mixed gas coming from mass loss from all of the bulge red-giants of different metallicities.
In this context, we note the recent results by Maness et al. (2007) who find for the very central part (within 1 pc) of the Galaxy
that their best fit model to constructed HR diagrams is consistent with continuous star formation over the last 12 Gyr 
with a top-heavy IMF. The similarity of this IMF to that IMF corresponding the most recent epoch of star formation derived
by Paumard et al. (2006) could suggest a possible connection between the stars that formed recently in the Galactic Center
with those that formed throughout its history.

\section{Conclusions}

Abundance patterns are discussed in two distinct populations: $\sim$ 12 Gyr old red-giants from the bulge 
and the young and massive M-type giants and supergiants residing mostly within 2 pcs of the central Black Hole. 
The abundance results indicate that the iron abundance distribution for the admittedly small sample of Galactic Center
targets studied so far shows very little abundance spread in contrast to the large metallicity spread found for
the old bulge population. Most importantly, both populations show enhancements in the $\alpha$-element abundances  relative
to the Galactic disk trend that might be explained invoking a top-heavy IMF.

\end{document}